\documentclass[10pt,letterpaper]{article}
\usepackage[top=0.85in,left=2.75in,footskip=0.75in]{geometry}

\usepackage{amsmath,amssymb}

\usepackage{changepage}

\usepackage[utf8x]{inputenc}

\usepackage{textcomp,marvosym}

\usepackage{cite}

\usepackage{nameref,hyperref}

\usepackage[right]{lineno}

\usepackage{microtype}
\DisableLigatures[f]{encoding = *, family = * }

\usepackage[table]{xcolor}

\usepackage{array}

\newcolumntype{+}{!{\vrule width 2pt}}

\newlength\savedwidth

\raggedright
\usepackage[aboveskip=1pt,labelfont=bf,labelsep=period,justification=raggedright,singlelinecheck=off]{caption}

\makeatletter
\renewcommand{\@biblabel}[1]{\quad#1.}
\makeatother

\usepackage{lastpage,fancyhdr,graphicx}
\usepackage{epstopdf}
\fancyheadoffset[L]{2.25in}
\fancyfootoffset[L]{2.25in}
\begin{document}
\vspace*{0.2in}

\begin{flushleft}
	{\Large
		%\textbf\newline{Science is Different: the Evolution of Competitiveness across Socio-economic Layers}% {Evolution of competitiveness in Science, Technology and Trade} % Please use "sentence case" for title and headings (capitalize only the first word in a title (or heading), the first word in a subtitle (or subheading), and any proper nouns).
		\textbf\newline{The Evolution of Competitiveness across Economic, Innovation and Knowledge production activities}
	}
	\newline
	\\
	Aurelio Patelli\textsuperscript{1*},
	Lorenzo Napolitano\textsuperscript{2,\ddag},
	Giulio Cimini\textsuperscript{3,1},
	Emanuele Pugliese\textsuperscript{2},
	Andrea Gabrielli\textsuperscript{1,4},
	\\
	\bigskip
	\textbf{1} Enrico Fermi Research Center, piazza del Viminale 1/a, Rome, Italy
	\\
	\textbf{2} Joint Research Centre (EC|JRC), Seville, Spain
	\\
	\textbf{3} Physics Department and INFN, University of Rome Tor Vergata, 00133 Rome, Italy
	\\
	\textbf{4} Engineering Department, University ``Roma Tre'', 00146 Rome (Italy)
	\\
	\bigskip
	
	% Insert additional author notes using the symbols described below. Insert symbol callouts after author names as necessary.
	% 
	% Remove or comment out the author notes below if they aren't used.
	%
	% Primary Equal Contribution Note
	% \Yinyang These authors contributed equally to this work.
	
	% Additional Equal Contribution Note
	% Also use this double-dagger symbol for special authorship notes, such as senior authorship.
	% \ddag These authors also contributed equally to this work.
	
	% Use the asterisk to denote corresponding authorship and provide email address in note below.
	* aurelio.patelli@cref.it

	\ddag The content of this article does not reflect the official opinion of the European Union. Responsibility for the information and views expressed therein lies entirely with the authors.

\end{flushleft}

\section*{Abstract}
The evolution of economic and innovation systems at the national scale is shaped by a complex dynamics, the footprint of which is the nested structure of the activities in which different countries are competitive.
Nestedness is a persistent feature across multiple kinds (layers) of activities related to the production of knowledge and goods: scientific research, technological innovation, industrial production and trade.
We observe that in the layers of innovation and trade the competitiveness of countries correlates unambiguously with their diversification, while the science layer displays some peculiar feature.
The evolution of scientific domains leads to an increasingly modular structure, in which the most developed nations become less competitive in the less advanced scientific domains, where they are replaced by the emerging countries. 
This observation is in line with a capability-based view of the evolution of economic systems, but with a slight twist. 
Indeed, while the accumulation of specific know-how and skills is a fundamental step towards development, resource constraints force countries to acquire competitiveness in the more complex research fields at the price of losing ground in more basic, albeit less visible (or more crowded), fields.
This tendency towards a relatively specialized basket of capabilities leads to a trade-off between the need to diversify in order to evolve and the need to allocate resources efficiently.
Collaborative patterns among developed nations reduce the necessity to be competitive in the less sophisticated fields, freeing resources for the more complex domains.
%\linenumbers

% Use "Eq" instead of "Equation" for equation citations.
\section*{Introduction}
%%% brief catch up
% 0) Evolution of the innovative systems is fundamental because...
The global set of national economic systems can be represented as a multitude of interrelated layers, each referring to a characteristic domain of activity (e.g. trade, innovation, scientific research), a subset of which are measurable.
Indeed, the view of economic outcomes as the result of the complex interactions between interconnected systems is a long-standing idea with deep roots e.g. in evolutionary economics. 
One notable example is the notion of systems of innovation~\cite{dr1987technology}, defined as the network of private and public institutions operating within a territory, which concur, through their activities and mutual relations, in the discovery an spread of new technologies. 
In general, the innovation system considers universities, firms and the public sector among the relevant actors responsible for the interactions that shape innovation and knowledge transmission. 
Moreover, the concept has been adapted successfully over time to serve as a framework to describe and analyze National~\cite{lundvall1992national, nelson1993national} as well as Regional systems~\cite{cooke2011handbook}.
A similar perspective is shared by the triple helix model~\cite{etzkowitz1995triple} and its later refinements, which identify economic agents and their co-evolution as drivers of knowledge production and innovation in knowledge-based societies, while reducing the emphasis on the importance of borders and local specificity.
As a result, understanding how the different layers evolve over time has become increasingly relevant in the economic literature and the variety of empirical tools proposed to address it has grown accordingly.
%TBC (looking for literature...)

% 1) 'Unfolding' groups the dynamics of tech,science and export -> coordination in their dynamics
The framework of Economic Complexity~\cite{Tacchella2012} allows to study the innovation systems focusing on the national competitiveness on the different activities.
The interaction between science, technology and production is rich and displays a high degree of interconnection.
Indeed, the strongest statistically significant signal of interaction between these layers suggests that technological breakthroughs drive the development of new products and science~\cite{Pugliese2017}, although the sub-leading interactions are still significant.

% 2) but we don't know how the single environment evolves
At the same time, each layer displays nested and diversified patterns with specific features that are not shared by the other layers.
These features are connected to the main drivers of competitiveness of each system and might not be general, although diversification arises in all the cases.
Therefore, it is interesting to measure the structural evolution of each layer taken in isolation and compare their analogies and differences.

% 3) how can we compare diverse systems? -> Nestedness and Modularity following ...
The scientific literature about biology and ecological systems provides a foundation for our analysis because mutualistic systems create patterns comparable to the ones found in economic systems.
Furthermore, the nested patterns of ecological systems are related to their stability with respect to external or internal perturbations.
Innovative and productive systems also display nested and modular patterns as emerging features.
Therefore, exploiting the analogy by applying the methods borrowed from the ecological domains could hep shed light on interesting properties of human systems, possibly related to their stability and evolution.

% 4) then, different results
The manuscript firstly introduces the material and methods, giving a ground to the datasets and the mathematical clues of the main tools considered.
Successively, the results are presented in detail and are argued in the final discussions.

\section*{Materials and Methods}\label{sec:methods}
This section describes the databases considered in the analysis and the tools and methods implemented.

\subsection*{Databases}\label{sec:data}
%%% OAG
The analysis of the scientific layer is based on data that aggregates the scientific output of nations into scientific domains.
The source of this data is the Open Academic Graph v2 (OAG)~\cite{MAG1,MAG2,MAKG} which is a snapshot of the Microsoft database taken in the late November 2018.
OAG lists a large number of academic papers, reporting for each one information such as its citation counts and the institute or university to which the authors are affiliated.
The database covers most of the journals, conference proceeding, books and manuscripts published from the early 1800s up to the moment when the snapshot was taken.
We consider publications starting from 1960 because in earlier years only a small number of publications is available, which mostly concentrate in a small set of developed countries.
In terms of geographical coverage, OAG accounts for most of the nations in the world and it is one of the most complete and detailed datasets in terms of the geographical coverage it allows.
OAG presents a small bias towards developed and English-speaking nations~\cite{Patelli2021geography}, although this is a common characteristic of many other databases, \textit{e.g.} Scopus~\cite{SCOPUS,Cimini2014}.

%%% REGPAT and COMTRADE
The technological dataset RegPat~\cite{regpat2020}, built by the OECD and updated yearly, accounts for the innovation taking place within a large set of countries as proxied by the patenting activity carried out by applicants and inventors located within their borders.
This database does not cover developed nations uniformly and, since it focuses on patent applications submited to the European Patent Office, it has a bias toward Europe.
Nevertheless, the database makes up for this shortcoming thanks to an accurate geocoding of the patent documents it contains. Overall, it contains data about 200 nations and 649 4-digits technological codes of the CPC classification. It covers the period from 1978, when the EPO was first established, to the year of publication.

The COMTRADE~\cite{COMTRADE} database collected by the UN, which reports the trade flows of physical goods between countries, forms the basis for the economic layer employed in our analysis.
The database, as homogenized by~\cite{Tacchella2018}, covers 169 nations and reports 1218 4-digits product codes of the HS-1992 classification.

\subsection*{Measures of competitiveness}\label{sec:measures}
%%% citation and log-citations
Technological impact can be measured with the number of patents filed by field of technology and economic impact can be measured with export flows by product category. Scientific impact is instead usually based on citation counts because citations are widely recognized as a proxy of the quality of the research performed by authors, institutions and, consequently, nations.
However, due to the rich-get-richer mechanism, citation counts display very skewed distributions with fat tails and weak convergence to stationary measures~\cite{Burrell2002,Radicchi2008,Eom2011}.
To partly correct skewness one can employ \textit{log-citation counts}~\cite{Fairclough2015,Medo2016,Patelli2021geography}, defined as
\begin{eqnarray}
	[lcit]_{i,\alpha} = \log(1+[cit]_{i,\alpha}).
\end{eqnarray}
where the label $i$ refers to geographical areas and the label $\alpha$ refers to the scientific field.
In the present manuscript we consider both citation counts and log-citations counts.
As the results show, the latter yield more stable results with lower fluctuations and are easier to interpret.

%%% RCA
Comparing countries based on their competitiveness requires identifying the domains actively pursued by each one.
This can be achieved using the Revealed Comparative Advantage (RCA) indicator~\cite{Balassa1965}, commonly used in literature as a measure of relative specialization.
As shown in equation~\ref{eq:rca} RCA is computed as the weight an activity in national baskets of activities relative to the global weight of the same activity.
\begin{equation}
	RCA_{i,\alpha} = \frac{W_{i\alpha}}{\sum_{\beta} W_{i\beta}} \Big/ \frac{\sum_j W_{j\alpha}}{\sum_{j\beta} W_{j\beta}}
\end{equation}\label{eq:rca}
where $W_{i,\alpha}$ indicates the extensive measure over which the RCA is computed, \textit{i.e.} patent counts, export flows or log-citations, depending on the layer.

%%% binary procedure
RCA takes values on a continuum and can hence encode a great deal of information.
However, for our purposes a more \textit{coarse grained} measure is more adequate. For this reason, we transform RCA into binary values flagging the activities in which each nation is more  \textit{active} then a given threshold.
Following standard practice, we apply binary filtering by keeping only RCA values above $RCA^*=1$, thus constructing binary bipartite networks whose adjacency matrix has elements $M_{ij}=\Theta(RCA_{ij}-1)$,  where $\Theta(\cdot)$ is the step function.

%%% Nestedness: Temperature and NODF
% A system is said to be nested when the elements that have a few items in them (locations with few species, species with few interactions) have a subset of the items of elements with more items
\subsection*{Nestedness}
Nestedness is a property of systems consisting of actors with heterogeneous features that measures the extent to which shared features belong to both feature-rich and feature-poor actors. 
In the manuscript we estimate the nestedness of each network through the computation of two metrics widely used in literature, the \textit{Temperature of Nestedness}~\cite{Atmar1993,Rodriguez-Girones2006} and the \textit{Nestedness metric based on Overlap and Decreasing Fill} (NODF)~\cite{Almeida-Neto2008}.
NODF estimates the nestedness evaluating the overlap of each row and column with respect to all the others rows and columns.
Defining $\mathbf{M}$ the bi-adjacency matrix of the network considered, NODF is computed as 
\begin{eqnarray}
	NODF(\mathbf{M}) = \frac{1}{\mathcal{N}}\left[\sum_{i,j}\theta(k_i-k_j)\frac{C_{i,j}}{k_j} + \sum_{\alpha,\beta}\theta(k_\alpha-k_\beta)\frac{C_{\alpha,\beta}}{k_\beta} \right]
\end{eqnarray}
where $\mathcal{N}$ is a suitable normalization and $C_{i,j}=\sum_\alpha M_{i,\alpha}M_{j,\alpha}$ is the number of co-occurrences between of rows $i$ and $j$, $C_{\alpha,\beta}=\sum_i M_{i,\alpha}M_{i,\beta}$ is the number of co-occurring element between columns $\alpha$ and $\beta$.
The function $\theta(x)$ is the Heaviside function, also known as step function.
The two terms inside the square brackets are proportional to the row NODF and the column NODF respectively.

On the contrary, the Temperature of Nestedness is computed through a rather convoluted formula evaluating the unexpectedness of 0/1 at the distance above/below the isocline coinciding with the line of perfect nestedness, which is determined by the density of the adjacency matrix.
We use the code made available by the Nestedness for Dummies (NeD) project~\footnote{\url{http://ecosoft.alwaysdata.net/}, project funded by the European Union} to compute the Nestedness Temperature.

\subsection*{Modularity}
Modularity measures the quality of the partitions of a network, i.e. a given community structure. According to the modularity metric, the best community structure maximizes
\begin{equation}\label{eq:mod}
	Q=\max_{\xi}\left\{\frac{1}{N}\sum_{i,j}\left(A_{ij}-\frac{k_ik_j}{2N}\right)\delta(\xi_i,\xi_j) \right\}
\end{equation}
where $\xi_i$ is the label of the partition to which node $i$ belongs and $A_{ij}$ is the adjacency matrix of the network.
In this work we compute the modularity of the monopartite projections of the bipartite networks corresponding to the layers connecting nations to their activities.
We focus mainly on the evolution of the modularity of each layer, where the elements of the adjacency matrix $A$ of equation~\eqref{eq:mod} are related to $C_{ij}$.
Moreover, we check that the modularity of the monopartite representation of each bipartite network, often considered in literature about the block nestedness, does not create meaningful partitions in terms of specialized blocks~\cite{Tur2015}.
The value of the modularity of the best partitioning measures the inter-dependence between the modules found, thus providing an estimate of the strength and stability of the proposed communities.

% Place figure captions after the first paragraph in which they are cited.
\section*{Results}
\label{sec:results}
\begin{figure}[!t]
	\centering
	\includegraphics[scale=0.275]{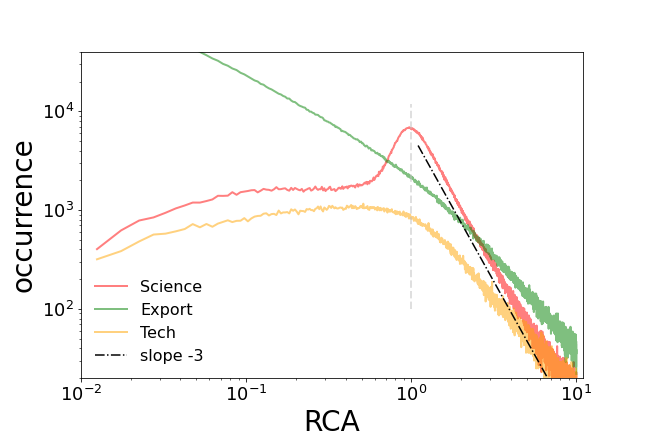}
	\includegraphics[scale=0.275]{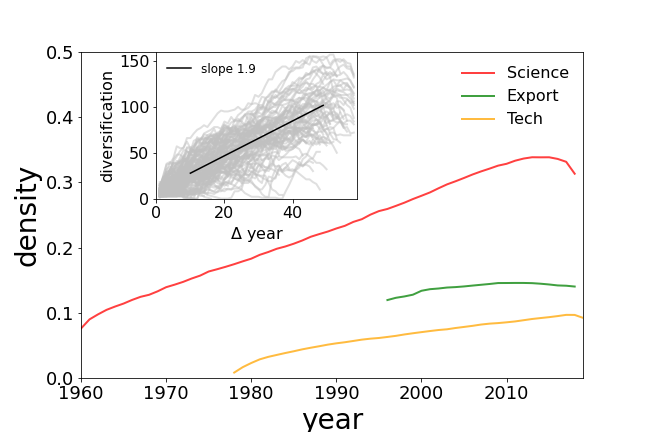}
	\caption{The left panel shows the occurrence of RCA for the different layers in log scale. 
		The dashed line indicates the position of the threshold value of $RCA=1$, while the full black line is an eye-guidance for a slope of $-3$.
		The right panel shows the density of elements having RCA larger than $1$ in the different layers considered: the Scientific (red line), the Technological (orange line) and the International Trade Export (green line).
		The inset indicates the grows of the scientific diversification of the different nations with respect to the initial time of production.}
	\label{fig:Distribution_RCA}
\end{figure}

%%% distribution of the RCA as a justification of the unitary threshold
As explained in the Materials and Methods section, the competitiveness of each nation in the various production and innovation activities can be estimated by evaluating the RCA.
The RCA of the economic layer is usually based on International Trade data that is considered a qualitatively good measure of competitiveness.
In the technological layer the RCA is often estimated by aggregating the patent production while in the scientific layer this is done through the log-citations counts.
Thus, the measure of RCA are different across layers (see figure~\ref{fig:Distribution_RCA}, right panel).
For example, the profile of the RCA distribution of the scientific data peaks around 1, with a lower occurrence of low RCA values and a stronger power-law decay corresponding to large RCA values.
Conversely, the economic data does not display a peak in the analyzed spectrum.
For all layers, the power-law decay is a sign that the system is highly heterogeneous since it cannot be easily replicated by the RCA distribution obtained by considering random matrices.
Yet, the slope of the decay depends on the particular properties associated with the relative competitiveness of the nations~\cite{Medan2007}.
Interestingly, the slope of the RCA distribution of the Technological and Scientific layers have a cross-over around 1 (the global average) while the distribution of the RCA of the international trade exports does not present a clear cross-over.

%%% density and the introduction to the triangularity
The binary representation of the RCA intrinsically defines the bipartite networks displaying the competitiveness of the nations in the basket of activities characterizing each layer.
A basic quantity of interest in this representation is the network density defined as the fraction of observed links with respect to the maximum number. \emph{i.e.} the number of links one would observe in the fully connected graph.
The time series of the density of the Scientific and Technological layers indicates that the nations increase their diversification as time increases, as shown in the right panel of figure~\ref{fig:Distribution_RCA}.
This growth marks a second difference of Science with respect to the Economy layer, which features a much steadier evolution of diversification.
Indeed, the density of the scientific environment grows almost linearly and it is probably triggered by the exponential growth of the scientific corpus with a doubling period of approximately a decade~\cite{Fortunato2018}.
Such exponential growth is not found only at the global scale but also at the national level for most of the developed and developing countries (as shown in the inset of the right panel of figure~\ref{fig:Distribution_RCA}).
A similar pattern can be observed in the production of patents, while it is not reproduced in the available Export data in the last decades, but it was detected during the economic boom around the sixties~\cite{Saracco2015}.

\begin{figure}[!t]
	\centering
	\includegraphics[scale=0.065]{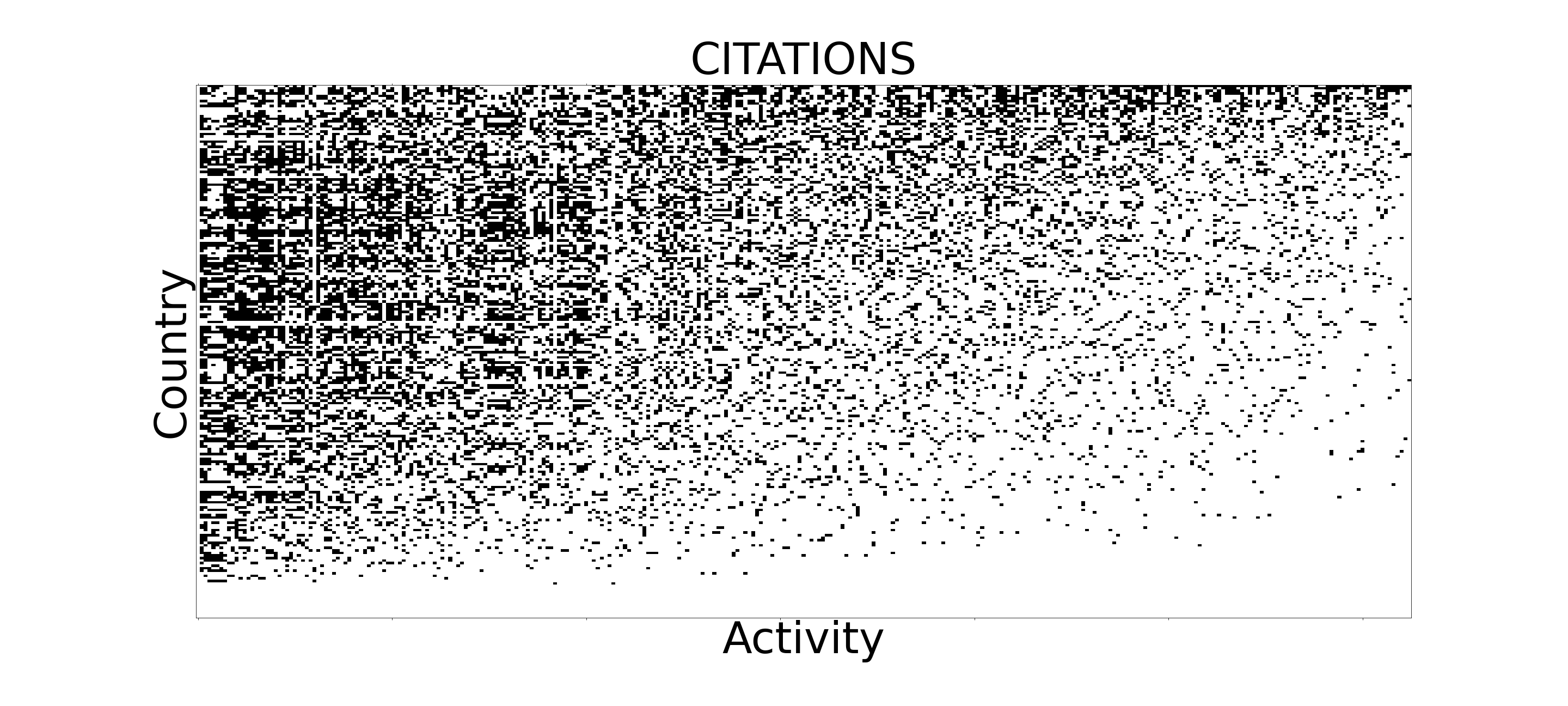}
	\includegraphics[scale=0.065]{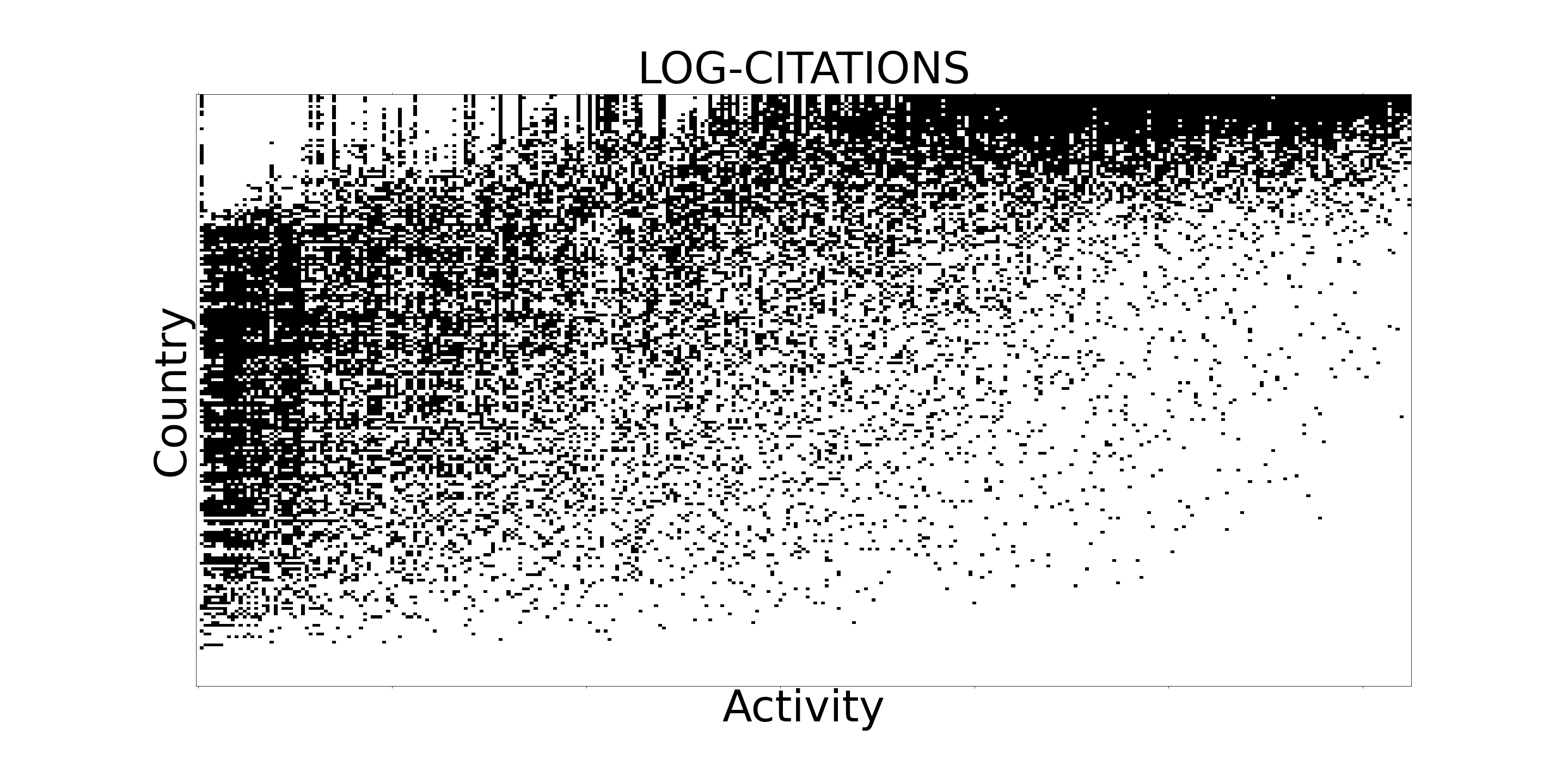}
	\includegraphics[scale=0.065]{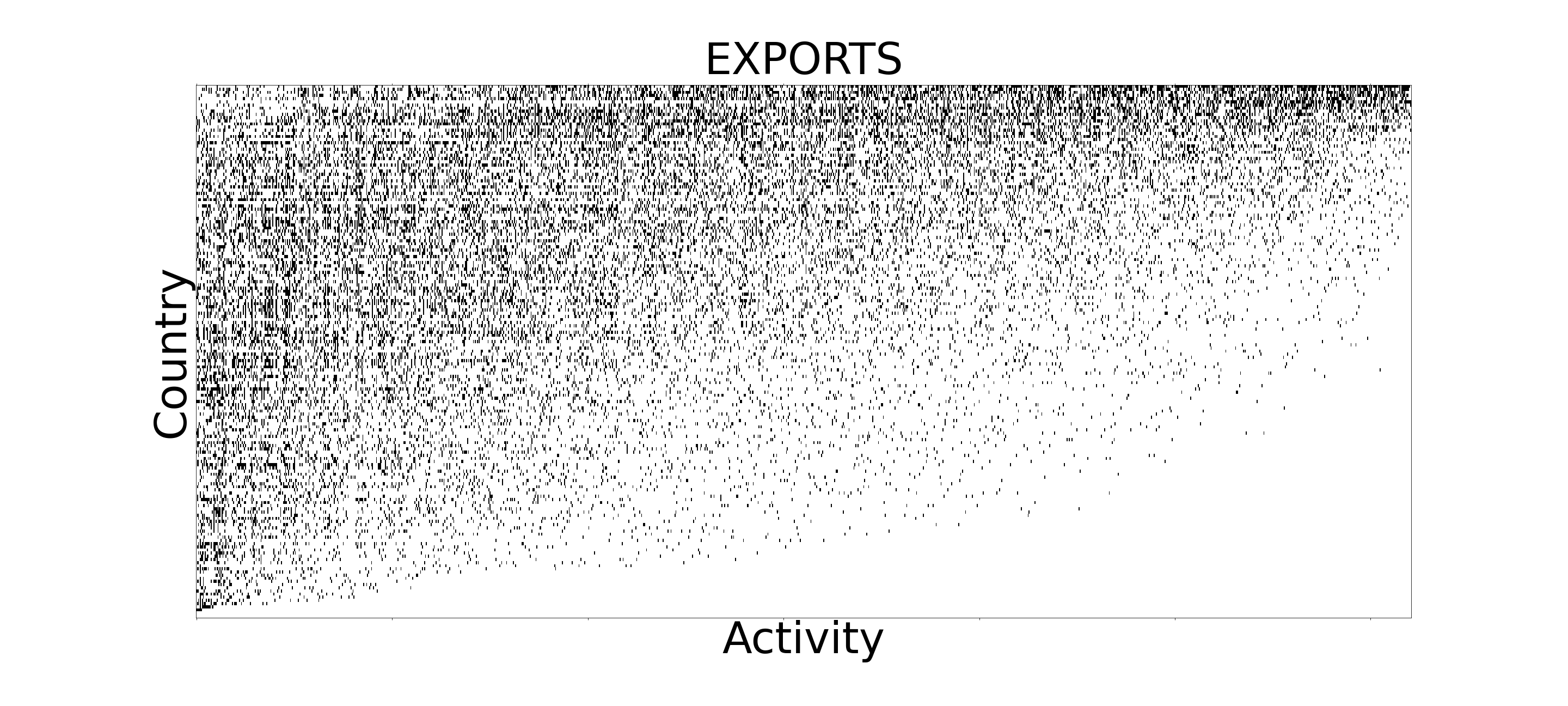}
	\includegraphics[scale=0.065]{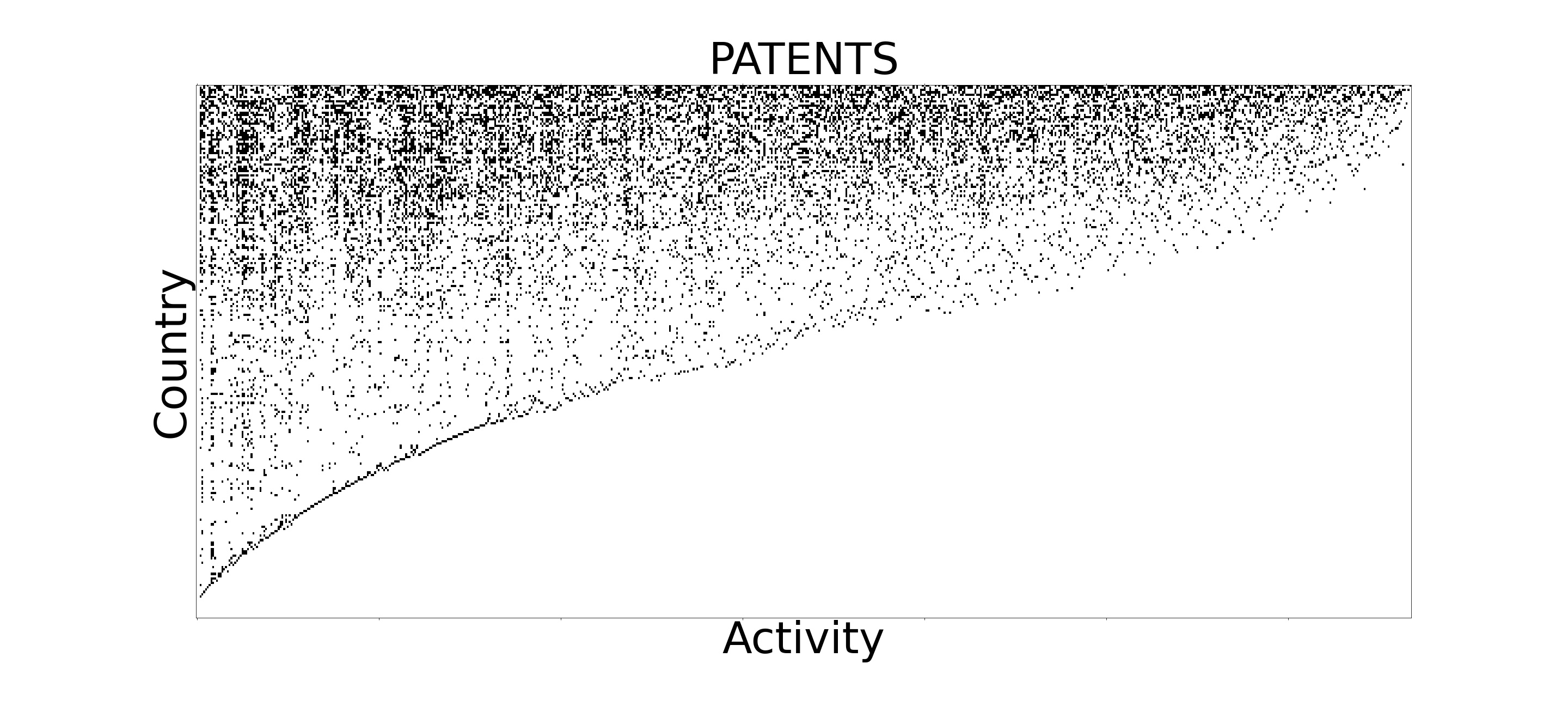}
	\caption{The binary matrices in $2000$ thresholding the RCA at 1 for based on the different networks: the scientific citation (top left), the log-citation (top right), the export flow (bottom left) and the patent production (bottom right).}
	\label{fig:Binary_MatricesRCA}
\end{figure}

%%% nestedness ...
The most important and characteristic pattern emerging from the binary representation is the presence of the triangular shape of the matrix $M_{i,\alpha}$, visible when the rows and columns are properly ordered~\cite{Hidalgo2009,Tacchella2012}.
All the layers display the hierarchical structure shown in figure~\ref{fig:Binary_MatricesRCA}, where top rows, corresponding to the most competitive nations, have a high diversification while some activities are highly ubiquitous (leftmost columns) and some others are performed only by the top nations (rightmost columns).
This feature of \textit{triangularity}, can be easily visualized when the matrices are properly ordered, and signals a high \textit{nestedness}, though, unfortunately, a precise mathematical definition of nestedness is still lacking~\cite{Mariani2019}.
Among the innovation-related layers, science appears to be the less nested, or triangular, while the technological layer displays a relatively sharp boundary along the diagonal, which highlights its nested structure nicely within the associated adjacency matrix .
This feature points to a structure of the scientific network that presents more intrinsic heterogeneity, compared to the other layers.
Such heterogeneity was considered as the cause of the lower quality of the scientific layer in the context of the Economic Complexity~\cite{Cimini2014,Patelli2017}, which was solved by the introduction of the more stable log-citations~\cite{Patelli2021geography} metrics.
Intriguingly, the network based on the log-citations metrics presents a hole in the top left corner, which is not observed in the other cases, suggesting that the top nations might be not very competitive in sectors with high ubiquity.
Note that the absence of the top left links does not mean that the top nations are not producing science in the less complex scientific domains, rather that the number of citations they receive in those fields is below the fair share, given the global average~\footnote{Indeed, looking at the value of the RCA in the hole region, the values are constantly close, but below, one.}.

%%% ... and its metrics
The lack of a precise definition of nestedness induces the derivation of different, albeit slightly counter-intuitive, metrics.
Indeed, depending on the feature considered as the representative characteristic of nestedness, various metrics can be defined to make the concept operational.
In this work we consider the Temperature of Nestedness~\cite{Atmar1993} (or, simply, Temperature), and the NODF~\cite{Almeida-Neto2008} because both are connected to different, yet related, features characterizing the dynamics of the innovative layers.
The main difference between the metrics is that Temperature estimates the nestedness by the unexpected presence/absence of links in the empirical bipartite network with respect to the perfectly nested case for a fixed density~\footnote{Remarkably, the fully nested structure implemented in the computation of the Temperature is a degenerate form where all the information can be obtained by the knowledge of the equilibrium shape, artificially setting the equilibrium diversification and ubiquity.}.
Instead, NODF estimates the nestedness by considering the degree of overlap that each row and column has with the others.
Hence, an important difference between the two metrics is that the algorithm computing the Temperature requires the network to be re-ordered to achieve the most nested arrangement, while NODF is independent on the ordering.
In the following we opt for the re-ordering given by the Fitness-Complexity ranks whenever we want to compute the Temperature, since the Fitness-Complexity algorithms has been shown to outperform other techniques~\cite{Lin2018} in approximating the maximally nested arrangement.

%%% simple interpretation of the measures of nestedness
Both measures of nestedness capture some feature of the dynamical evolution of the bipartite networks, and consequently of the innovation systems.
For instance, by computing the overlap  (co-occurrence) between nations, NODF can describe the ability of countries to follow the path of more developed nations, according to capability-based mechanisms.
Indeed, NODF can be separated into the row and column components, allowing to disentangle the contribution of row-wise and column-wise co-occurrences to the nestedness of the system. %separating the co-occurrence on the respective layers.
On the contrary, Temperature evaluates the match and difference of the empirical network with a fully nested equilibrium of the environment, based on the stable state of a mutualistic system~\cite{Medan2007}.
Therefore in the innovation systems, the Temperature is high when low performing nations are active in highly unexpected domains.% which may be induced by the presence of correlations among the natural progression of activities in time.

%%% necessity of a statistical validation
A problem encountered in the evaluation of the nestedness is that Temperature and NODF may depend on more basic topological properties of the networks that are not related to a particular visual pattern.
For instance, the density of the network is the most important parameter in the estimation of the nestedness~\cite{Mariani2019}, so that comparing networks with different densities is problematic.
Another typical source of bias in the comparison of the nestedness is given by the degree distributions (the distribution of the diversification of countries and of the ubiquity of activities) since their evolution is not random and presents high temporal persistence.
The standard way to correct this issue is to extract the statistical significance of the nestedness, scattering the empirical measures with respect to those obtained in suitable random models able to represent the selected biases~\cite{Cimini2019,Cimini2022}.
For instance, the Erdos-Renyi (ER) null model~\cite{Newman2010} draws a network ensemble constraining only the average density, while the Bipartite Configuration Model (BiCM)~\cite{Saracco2015} constrains also the average network degrees.

\begin{figure}[!t]
	\centering
	\includegraphics[scale=0.3]{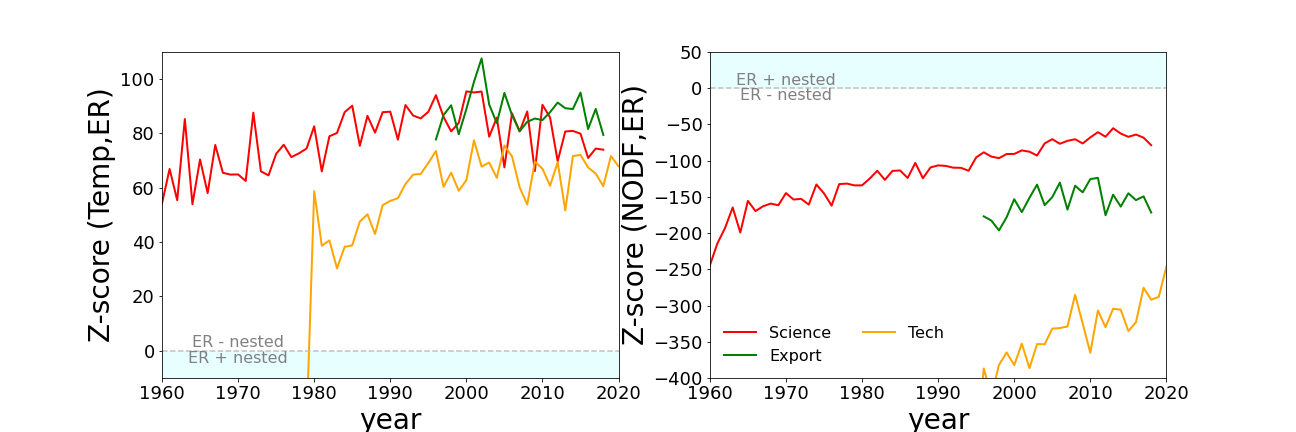}
	\includegraphics[scale=0.3]{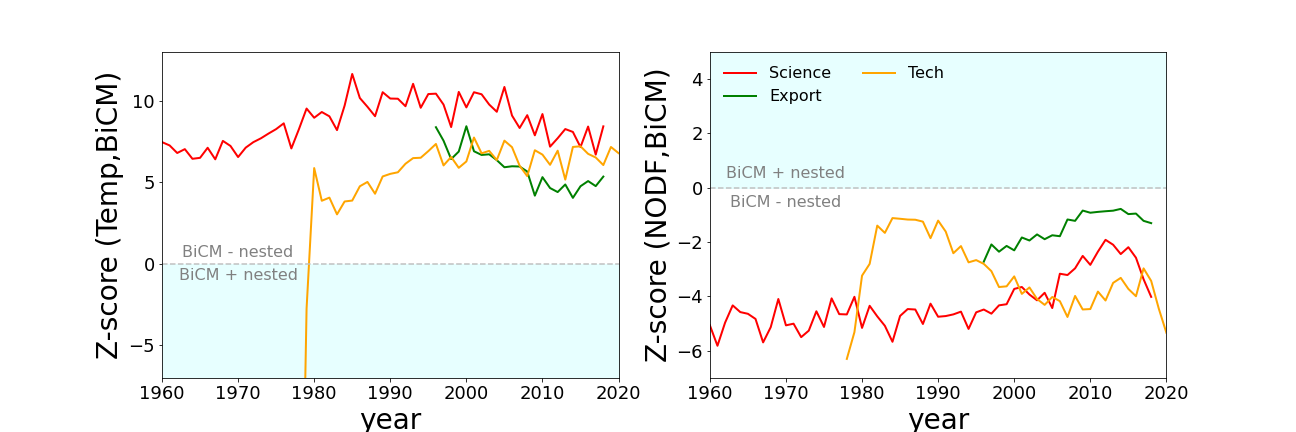}
	\includegraphics[scale=0.3]{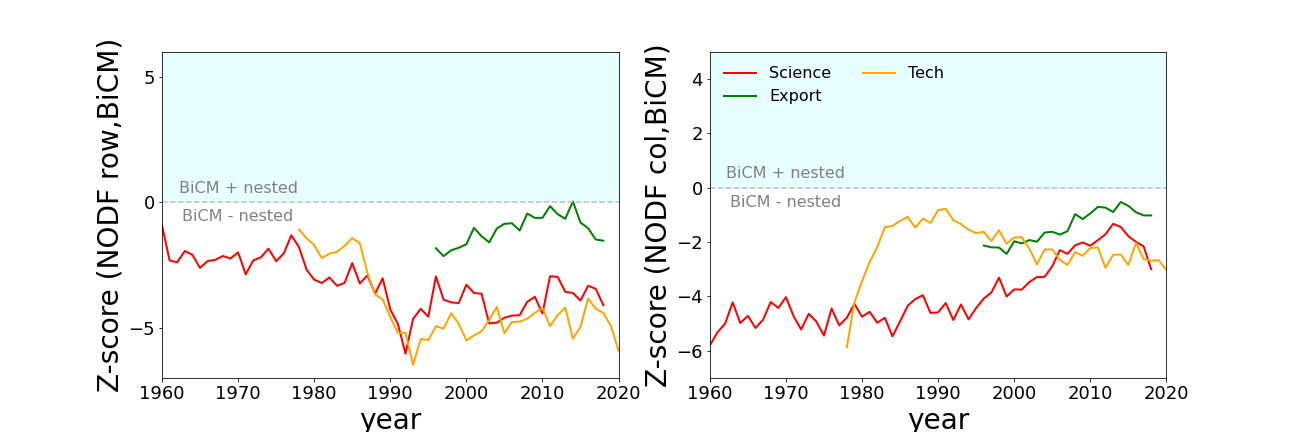}
	\caption{Temporal evolution of the Z-scores for the empirical measurements of nestedness (either Temperature or NODF) against a null network model (either ER or BiCM, which respectively constrain density and degree distributions). The different lines correspond to the various layers: Science (red), Technology (orange) and Trade (green). The light blue area marks the region of the plot where the null model is more nested than empirical data. The difference on the positions of the cyan regions derives from the inverted ranges of Temperature and NODF.}
	\label{fig:temperature_ER_BiCM}
\end{figure}

%%% ER does not extract much information
Irrespective of the increasing trend of both the density and the nestedness measures over time, discounting only the density does not provide much information in terms of evolution of the nestedness in the innovations systems.
Indeed, the ER ensemble of matrices is much less nested than the empirical matrices, as shown in the top panels of figure~\ref{fig:temperature_ER_BiCM}.
According to the z-score values, which measures the distance of empirical values from ensemble averages in terms of ensemble standard deviations, in all layers empirical nestedness is much higher than in the random case.
Instead, the null ensemble obtained by constraining the degrees of the nodes in the  network with the BiCM, leads to a much lower significance of the features of the empirical network.
Indeed, the value of the degrees is not random but has a persistent dynamics that affects the evolution of the nestedness, and this information must be accounted for in the null models.
The information contained in the degrees is not usually taken into account in the biological literature on nestedness, where most of metrics were originally developed~\cite{Bastolla2009} (see however \cite{Borras2019,Bruno2020}).
For example, the perfect patterns for the Temperature are related only to a single parameter, the density.
On the contrary, in the Economic literature, diversification (\emph{i.e.} the node degree in the binary network) is the main feature against which the nestedness is studied~\cite{Cristelli2015}~\footnote{However, temporal series exploiting the dynamics of the systems can be obtained more easily in the Economic literature, thus, it is possible that also in the Ecological cases diversification could be important.}.
However, the difference between the significance of Temperature and NODF is not strong and for what concerns this analysis, the two nestedness measures correlate.

%%% modularity
The nested pattern that emerges in innovation layers is usually related to a competitive dynamics.
Instead, collaborative systems can create a more clustered structure of the networks, with the growth of modules or communities~\cite{SoleRibalta}.
In the biological realm, the modular organization of species interactions can increase the dynamical stability of the communities~\cite{Grilli2016,Sheykhali2020} toward exogenous (external) perturbations~\footnote{Instead, negative modularity might destabilize ecological systems.}.
The combination of competitive and collaborative dynamics promotes the formation of local nested patterns or in-block nestedness, indicating the natural aggregation of components on the network.
\begin{figure}[!t]
	\centering
	\includegraphics[scale=0.3]{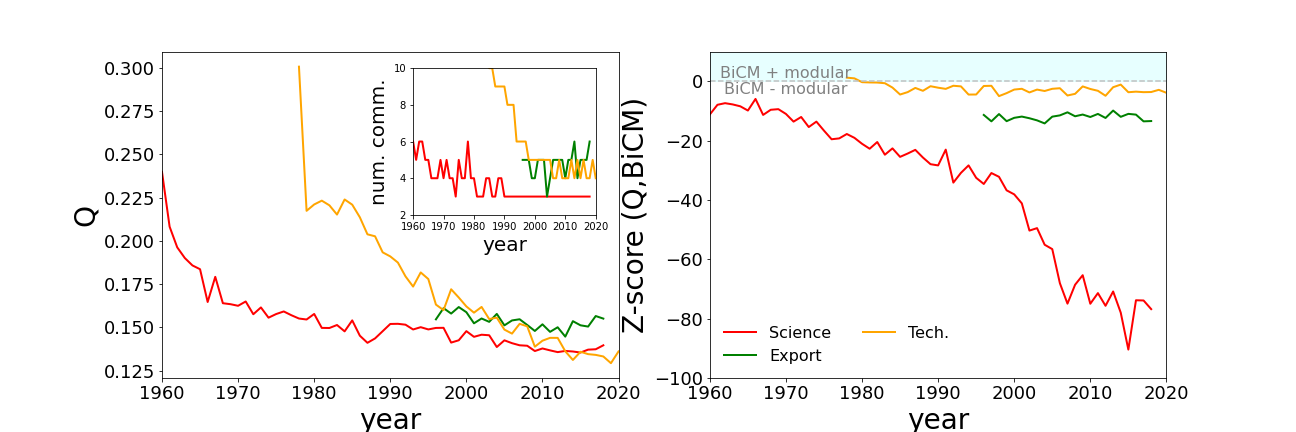}
	\caption{Left panel: the temporal evolution of the best modularity of the nations projection within the different layers.
		The inset indicates the corresponding number of communities.
		Right panel: the temporal evolution of the z-scores of the modularity of the different layers. 
		The line correspond to the Scientific (red line), the Technological (orange line) and the International Trade Export (green line).
		The light blue region on the diagram indicates the region where BiCM is more modular.
	}
	\label{fig:modularity}
\end{figure}
Focusing on the modular evolution of communities of nations having many internal connections (resulting from co-occurrences of activities) than connections with other communities, all the layers tend to level out at a comparable value of the optimal modularity, as seen in the left panel of figure~\ref{fig:modularity}.
The decrease of the value of the best modularity is typically associated to a decrease in the quality of the modular structures~\cite{Grilli2016,Palazzi2019} due to a more uniform distribution of links within and across partitions. %link between inmates and foreign nations.
However, the effective number of modules, or communities, seems to converge to a number between 3 and 5, suggesting that the network is getting more and more clustered, as a result of a more globalized and collaborative world.
Interestingly, the best partitioning of the scientific activities indicates neither a clear organization nor a block-nested structure.
Indeed, only on the layer of nations a meaningful partition is found while the scientific fields are more homogeneously distributed, thus the emergent community structure is dependent on how different classifications separate the fields of science. %their possible organization strongly depends on their classification.

Following the same arguments discussed above regarding the nestedness, the modularity depends on the network's features and the comparison of different networks is performed using the random model for the bias-removal.
The right panel of figure~\ref{fig:modularity} indicates that the modularity in science becomes more and more significant, thus the partitions are more and more reliable.
On the contrary, both the international trade export and the technological layers are closer to a random structure, indicating that the modules are less meaningful and can be partially ascribed to the diversification and ubiquity.

\section*{Discussion}
%%% few generalities
In this Section we discuss possible interpretations of the results.
%%% interpretation applied to the cases (science)
%%% diversification from the statistical validation
The evolution of the scientific system highlights a constant growth of diversification: all countries are enriching their basket of active domains.
Such growth persists from the end of the second world war.
The less competitive nations become able to actively progress in more sophisticated research areas following the leading countries, while the leaders are developing new fields of research increasing their diversification further.
This picture is consistent with the standard narrative of the evolution of innovative systems, from the Economic to the Technological layers, where the evolution of the competitiveness is related to the evolution of capabilities brought by each single actor and nation~\cite{Dosi2000,hausmann2007you,Bustos2012}.
The same conclusion can be drawn here because the most reliable patterns can be obtained by the null model built by fixing the diversification and ubiquity on the empirical values (see figure~\ref{fig:temperature_ER_BiCM}).
Instead, fixing only the global density generates an ensemble where the average behavior is very far from the evolution of the empirical patterns.

%%% problem with citations
%Following the capability-based narrative, the evolution of the scientific environment derives from the reference measure of competitiveness in science, the RCA based on the citations counts, where the less performing nations become more and more competitive.
%However, the intrinsic noise brought by the citation counts reduces the predictive power of the metric because the rich-get-richer mechanism, typical of the citation dynamics~\cite{}, may enhance the noise strength of the metric and create biases in the countries with low capabilities.
%Actually, parallel to the increase of the global density and of the number of competitive nations, also the significance of the citations-network nestedness reduces linearly in time with both the Temperature and the NODF estimations, as seen in the central panels of figure~\ref{fig:temperature_ER_BiCM}.

%%% the log-citation
%An effortless way to reduce the intrinsic noise of the scientific environment discussed above is the computation of the scientific bipartite network based on the log-citations, which shrink the RCA value around the global average, as shown in figure~\ref{fig:Distribution_RCA}.
%This methodology permits to obtain a more stable estimation of the scientific competitiveness of the nations, both in terms of persistence and of stability of the Fitness values~\cite{}.
Furthermore, the nestedness of the network remains stable in the significant region considering both the Temperature and the NODF estimations, as seen in the central panels of figure~\ref{fig:temperature_ER_BiCM}.
However, the binary matrix representation of the network indicates a new pattern hidden from the implementation of the citation counts metrics: two denser regions appear on the top right and bottom left sides with a hole in the top left corner.
The appearance of two regions in the country layer suggests the emergence of a modular organization of nations that cannot be described by the sole dynamics of diversification, but needs to account a specialization mechanism.
Thus, the region on top is populated by the most diversified and competitive nations that are focusing on the most complex scientific domains, reducing their competitiveness in the less complex sectors.
Instead, the less competitive nations have more probability to be competitive in less sophisticated domains and are less efficient in the more complex sectors, as indicated by the fact that according to Temperature the empirical network is significantly more nested than the random.

%%% interpretation
A possible interpretation is that the evolution of the top nations brings them to dismiss resources allocated from the less complex fields favoring the most complex domains, which causes the appearance of the hole in the top left corner that is not present in the standard narrative of diversification~\cite{Dosi2000}.
This hole is characterized by RCA values below the global average, although these values remain close but below the threshold.
Thus, in the scientific environment the driver of evolution is not simply the effort to achieve an increasing diversification of the basket of activities but, rather, a trade-off strategy between diversification and specialization in order to better allocate the available resources.
This behavior is probably the trademark of the scientific layer, where funding is usually channelled toward the most complex and highly-cited domains, instead of being broadly distributed throughout the spectrum of scientific activities.
For instance, in the production of physical goods there is an economic advantage to produce also the less complex artifacts, while this feature is less important in the scientific realm.
At the same time, the scientific evolution of the less performing nations is driven by emulation of the dynamics of leading nations with an increase of diversification, as indicated by the high significance of NODF.
However, the significance of the Temperature suggests that, although the less performing nations are able to be active in some sophisticated domain, this is lower that the random expectation.

%%% something about the technological environment...
The Technological layer yields the visually best nestedness because the plot of the network's adjacency matrix presents a clear triangular shape.
The most diversifies nations lie at the top of the image with a roughly uniform basket of active sectors, ranging from the less to the most complex domains.
The presence of a clear diversification among the nations suggest that the dynamics of the Technological environment follows the capability-based framework and for instance, NODF becomes more significant over time.
Indeed, the NODF of rows (countries) follows that of the Scientific layer and the difference between the two relies on the different activities layers.
Furthermore, the significance of the Temperature follows the same trend since the significant level of unexpectedness is comparable.
Instead, the Technological environment is the less prone to create meaningful partitions, or modules, among the nations, and its dynamics is probably driven by the co-occurrence of knowledge and capabilities.

%%% economic
Finally, the Economic environment displays a more conservative evolution, keeping roughly constant the network density and the effective number of active competitors in the system.
Thus, the dynamics of the network is more related to an evolution of the nestedness pattern and not of its size.
Indeed, the nestedness follows the dynamics depicted by the other Technological systems, highlighting the possible presence of common features of the dynamics of innovations.

\nolinenumbers

\bibliographystyle{plos2015}
\bibliography{main.bib}

\end{document}